\documentclass[journal]{IEEEtran}
\usepackage{times}
\usepackage{graphicx}
\usepackage{amsmath}
\usepackage{amssymb}
\usepackage[T1]{fontenc}
\usepackage{graphicx}    
\usepackage{subcaption}  
\usepackage{booktabs}
\usepackage{array}   
\usepackage{xcolor}  
\usepackage{multirow}

\title{Toward Low-Altitude Airspace Management and UAV Operations: Requirements, Architecture and Enabling Technologies}
\author{Guiyang Luo, Jinglin Li, Qixun Zhang, Zhiyong Feng, \\ Quan Yuan, Yijing Lin, Hui Zhang, Nan Cheng, and Ping Zhang

\thanks{Guiyang Luo, Jinglin Li, Quan Yuan, Yijing Lin, and Ping Zhang are with the State Key Laboratory of Networking and Switching Technology, Beijing University of Posts and Telecommunications, Beijing 100876, China (e-mail: luoguiyang@bupt.edu.cn; jlli@bupt.edu.cn; yuanquan@bupt.edu.cn; yjlin@bupt.edu.cn; pzhang@bupt.edu.cn) (corresponding author: Jinglin Li).}

\thanks{Qixun Zhang, and Zhiyong Feng are with the Key Laboratory of Universal Wireless Communications, Ministry of Education, Beijing University of Posts and Telecommunications, Beijing 100876, China (e-mail: zhangqixun@bupt.edu.cn; fengzy@bupt.edu.cn).}

\thanks{Hui Zhang is with the School of Computer and Information Tech-
nology, Beijing Jiaotong University, Beijing 100876, China (e-mail:
huizhang1@bjtu.edu.cn).}

\thanks{Nan Cheng is with the Key State Laboratory of ISN and the School of
Telecommunications Engineering, Xidian University, Xi'an 710071, China
(e-mail: nancheng@xidian.edu.cn).}
}

\begin{document}

\maketitle

The low-altitude economy (LAE) is rapidly advancing toward intelligence, connectivity, and coordination, bringing new challenges in dynamic airspace management, unmanned aerial vehicle (UAV) operation, and security management. Existing systems remain fragmented and lack effective coordination. 
To bridge these gaps, we propose UTICN (Ubiquitous and Trusted Intelligent Cellular-native Network) for LAE, a unified cellular-native architecture that integrates multi-domain sensing, high-precision positioning, intelligent aircraft-to-everything communication, dynamic airspace management, and UAV operational services. UTICN introduces key technologies such as integrated sensing and communication (ISAC), passive and active positioning, intelligent machine communication, swarm coordination, and control-data decoupled management frameworks. 
We demonstrate UTICN's feasibility through two use cases, i.e., a city-level LAE management platform and a multi-frequency collaborative ISAC system. This work provides a fundamental reference for building a unified operational foundation and airspace management architecture for the LAE. 


\section{Introduction}

The low-altitude economy (LAE) is emerging as a pivotal strategic domain for the future utilization of airspace resources. A cornerstone of this advancement is the development of LAE networks (LAENets), which should be a ubiquitous, intelligent, and interoperable network infrastructure capable of seamlessly connecting unmanned aerial vehicles (UAVs) with terrestrial digital ecosystems \cite{2025arXiv250421583W}, as well as providing airspace and security management. LAENets are crucial in supporting a diverse range of mission-critical applications, including urban logistics, infrastructure monitoring, aerial surveillance, electrically powered vertical takeoff and landing aircraft, and emergency rescue \cite{10388419}. 

These applications impose stringent requirements on dynamic airspace management, UAV operational service, and safety assurance \cite{zaid2023evtol}.  To meet these demands, LAENets must offer persistent spatiotemporal environmental sensing, high-precision positioning, mission-critical communication, and scalable cross-domain management and service capabilities. However, current architectures fall short due to fragmented design, limited sensing capabilities, and insufficient cross-system coordination. For example, small UAVs are difficult to detect in dense urban environments; coordination across flight zones or jurisdictions remains inefficient; and existing systems lack timely response mechanisms for anomalies or security events \cite{3gpp_ts_23256_v19_1_0}.

To meet these growing demands, the LAE is evolving along three key technological trajectories. First, intelligence enables UAVs to perceive and make decisions through onboard artificial intelligence (AI). Second, connectivity is enabled through a combination of wide-area air-ground integration and heterogeneous technologies, including 5G-advanced (5G-A), reduced capability (RedCap), and aircraft-to-everything (A2X) communication \cite{3gpp_ts_24577_v19_0_0}. Third, effective system-level collaboration necessitates real-time coordination between UAV swarms, airspace infrastructures, and key stakeholders, including government authorities, UAV pilots, and air navigation service providers. These converging trends provide the essential basis for the development of large-scale, ubiquitous, trusted, controllable, and operational LAENets.

Multiple research efforts have explored over LAENets. Telecom operators have proposed integrated sensing and communication (ISAC) architectures based on 5G-A~\cite{10929156}; aerospace and geospatial communities focus on fusing radar for high-precision positioning; and air navigation service providers promote predefined airspace grids and structured routing \cite{CHEN2024102573}. While effective in specific contexts, these efforts often operate in silos, targeting either sensing, communication, or routing, without a unified, holistic framework that integrates sensing, connectivity, localization, management and service. This fragmentation limits their applicability in complex, large-scale UAV operations and airspace management.

To fill this gap, we propose a Ubiquitous and Trusted Intelligent Cellular-native Network (UTICN) for the LAE. UTICN adopts a unified cellular-native architecture that integrates sensing, positioning, communication, management, and service to provide support for safe, ubiquitous, trusted, controllable, and operational LAE. It features key enabling technologies such as ISAC, active and passive positioning, air-ground intelligent machine communication, control-data decoupled management, and swarm intelligence-driven situational evolution. Our proposed solution is validated through two representative deployments: a city-scale LAE management system and a cooperative multi-frequency ISAC system. 

The remainder of this paper is organized as follows. Section \ref{sec-II} outlines the key requirements for UTICN. Sections \ref{sec-III} and \ref{sec-IV} introduce the UTICN architecture and elaborate on its core components and enabling technologies. Section \ref{sec-v} presents two typical application cases. Finally, Section \ref{sec-VI} concludes this work.

\section{LAE Requirements} \label{sec-II}
\begin{figure*}[t]
\centering
\includegraphics[width=0.88\linewidth]{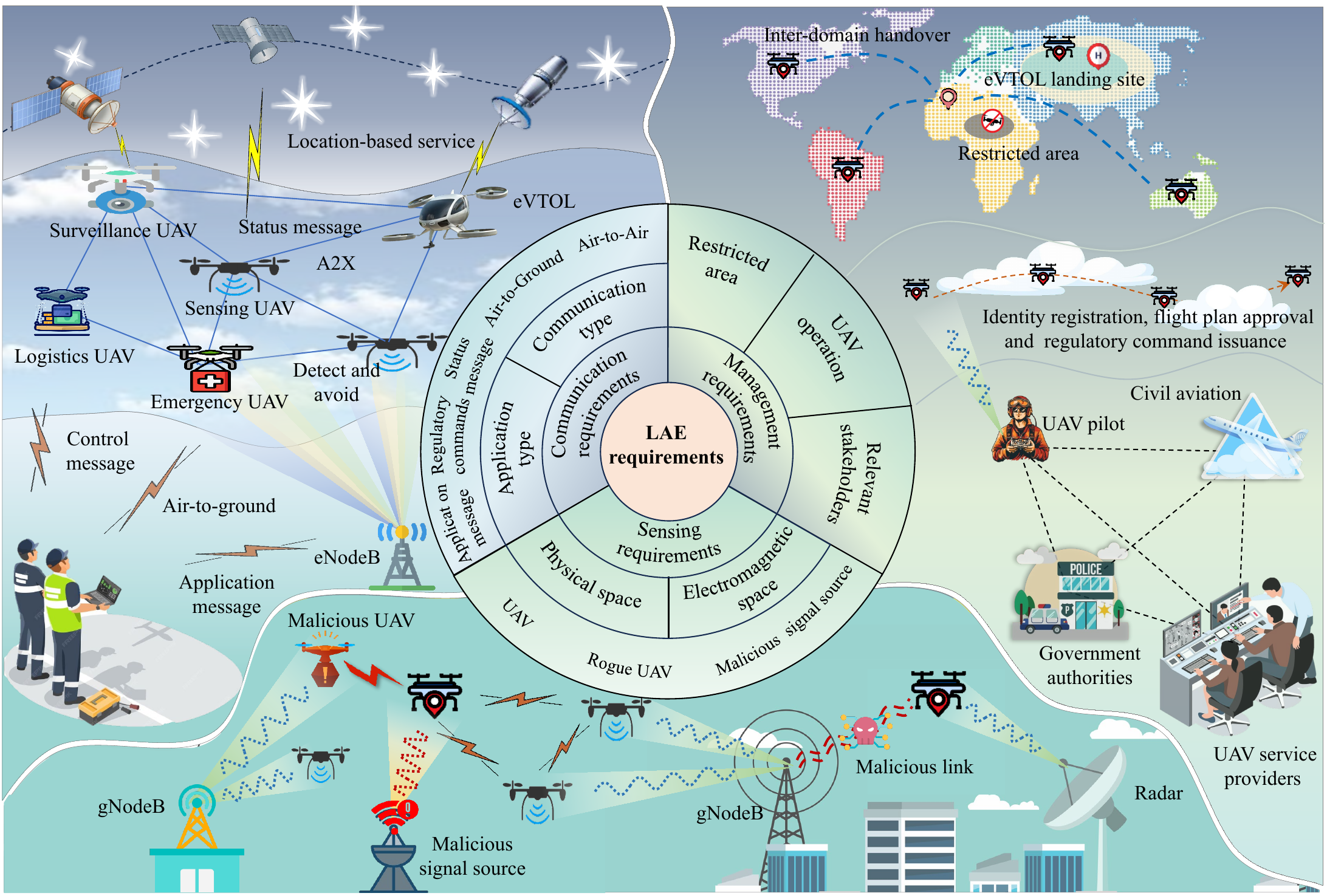}
\caption{LAE requirements, including sensing, communication, and management requirements. }
\label{fig:needs}
\end{figure*}
LAE demands the development of a ubiquitous and trusted infrastructure that not only facilitates interconnection among UAVs, relevant stakeholders, e.g., government authorities, UAV service providers, and air navigation service providers, but also ensures global observability and governance, persistent manageability, and timely controllability of low-altitude operations \cite{he2025satellite}. 

Despite increasing UAV activity, LAE systems still suffer from foundational challenges summarized as invisible, disconnected, and uncontrollable. These issues stem from the absence of a unified and holistic framework that spans the full lifecycle, operational chain, and stakeholders. Currently, UAVs predominantly rely on isolated point-to-point communication links, executing pre-planned missions with limited real-time environmental awareness and without dynamic access to airspace constraint information. As a result, most systems cannot reliably identify abnormal targets, trace their sources, or respond effectively to unauthorized behaviors, accumulating substantial operational risks.

To enable safe and orderly UAV operations and airspace management across diverse scenarios, it is essential to establish a unified and holistic platform that provides dynamic airspace management, unmanned aerial vehicle operation, and safety management. The platform must support the full process, from mission approval and identity registration to task execution, while ensuring seamless cross-domain networking, dynamic link handover, and system interoperability ~\cite{jiang20236g}. Moreover, it must enable timely responses and effective control of abnormal and unsafe events, such as unauthorized UAV activities (e.g., rogue drones) \cite{li2024unauthorized}. Meeting these goals requires that the platform satisfy three core capabilities:

\textbf{Sensing requirements}: multi-dimensional real-time awareness spanning both the physical and the electromagnetic space. In the physical space, the sensing system must deliver continuous, wide-area, and fine-grained coverage, capable of operating in complex urban settings with dense buildings and dynamic weather conditions. It should support all-weather, all-domain monitoring, enabling accurate identification, classification, and tracking of both legitimate and rogue UAVs. Furthermore, region-level situational awareness is essential for monitoring high-risk or sensitive areas, capturing traffic density, trajectory deviations, and emergent threats to support large-scale, coordinated UAV scheduling. In the electromagnetic space, the sensing infrastructure must provide real-time visibility into the spectral landscape of the low-altitude airspace. This includes monitoring authorized communication links, identifying abnormal or malicious signals, and analyzing electromagnetic behaviors across frequency, time, and spatial dimensions. Robust electromagnetic sensing is critical to ensure reliable communications, detect unauthorized transmissions, and support timely safety interventions and public security enforcement.
 
\textbf{Communication requirements}: high reachability, stability, and mission-critical connectivity. With the increasing number and intelligence level of UAVs, the communication system must go beyond basic support for UAV control and task data transmission. It should enable efficient state exchange and coordination among UAVs, as well as among UAVs, LAE infrastructures, and relevant stakeholders. Especially in complex urban environments, communication links must maintain high accessibility and robustness, ensuring UAVs can continuously receive regulatory commands, upload operational states, and promptly respond to airspace scheduling and emergencies.
Moreover, the system should support diverse communication modalities such as air-to-air and air-to-ground links, enabling unified access and hierarchical management across different tiers and types of airspace participants. It must also incorporate differentiated protection mechanisms for control signaling and mission payloads, ensuring higher priority and reliability for regulatory instructions.

\textbf{Management requirements}: LAENets should support full-process management from identity registration and flight plan approval to real-time airspace monitoring and command issuance. It must enable rapid identification and intervention of abnormal behaviors and allow coordinated regulation across government authorities, UAV pilots, and air navigation service providers. To build a sustainable and scalable low-altitude regulatory framework, the system must fulfill three essential capabilities: airspace observability to monitor flight environments, operational status, and abnormal targets; operational manageability to support approval, scheduling, alerts, and trajectory guidance; and behavioral controllability to deliver commands and enforce interventions against illegal or unsafe operations.

In summary, building LAENets is a systematic enhancement of UAV regulation, UAV operational services, airspace management, and public safety, forming the digital backbone of the future LAE.

\section{UTICN Architecture} \label{sec-III}

\begin{figure*}[t]
\centering
\includegraphics[width=0.8\linewidth]{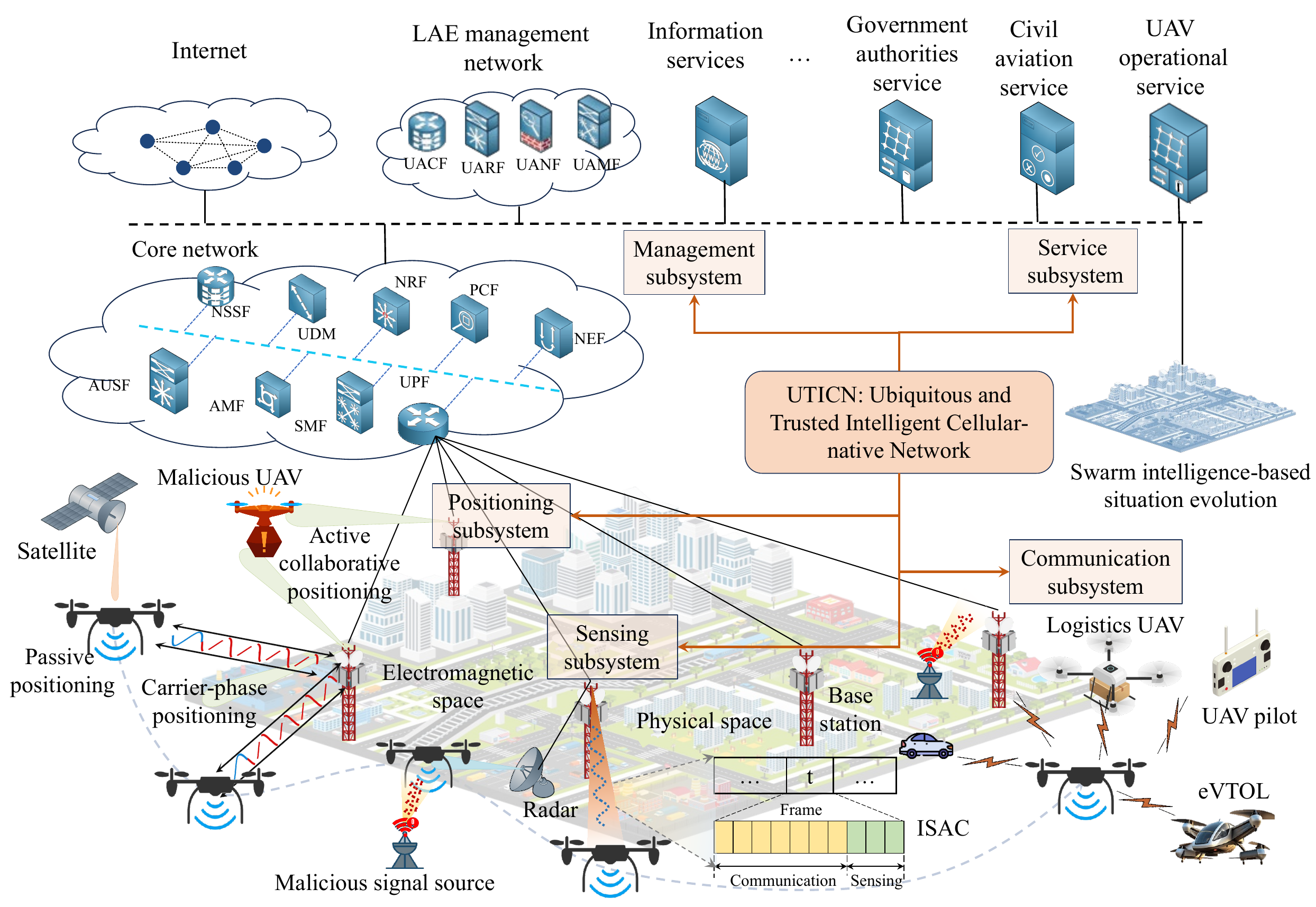}
\caption{UTICN, a ubiquitous and trusted intelligent cellular-native network, which integrates five core functional subsystems, i.e., sensing, positioning, communication, management, and service subsystem. It provides a systematic solution for UAV regulation, UAV operational services, airspace management, and public security management.}
\label{fig:UTICN}
\end{figure*}

To fulfill these requirements with an integrated cellular-native network, we propose UTICN, a unified cellular-native architecture designed to seamlessly integrate the physical and digital dimensions of low-altitude ecosystems. As illustrated in Fig. \ref{fig:UTICN}, UTICN integrates five core functional subsystems, i.e., sensing, positioning, communication, management, and service, into a unified, biquitous, trusted, controllable, and operational infrastructure for next-generation airspace management and UAV operations. 

\textbf{Sensing subsystem}: Enables multi-domain situational awareness by integrating multi-modal sensing technologies over both the physical space and the electromagnetic space, including radar, infrared vision, electromagnetic sensing, A2X communication, and ISAC. ISAC, a key enabler in 6G systems, utilizes embedded reference signals \cite{wei20245gdmrsbasedsignalintegrated} to support real-time sensing over UAVs, airspace, and infrastructures. In addition, electromagnetic environment sensing enables interference localization and signal integrity monitoring, ensuring robust support for both navigation and communications. 

\textbf{Positioning subsystem}: Provides high-precision and continuous localization by integrating passive, e.g., Global Navigation Satellite System (GNSS) and real-time kinematic differential positioning,  and active, e.g., cellular carrier-phase methods, and radar/vision-based techniques, positioning technologies. Cellular carrier-phase techniques deliver centimeter-level accuracy in GNSS-challenged environments (e.g., tunnels or dense urban areas). Through multi-source fusion, the system refines coarse estimates and enables accurate, reliable positioning essential for UAV operations and control in complex scenarios.

\textbf{Communication subsystem}: Supports wide-area and mission-critical air-ground and air-air communications, which are critical for UAV operation, airspace management, data transmission, and cooperative mission execution. It integrates dedicated data links, 5G RedCap communication, PC5-based A2X direct connectivity \cite{jiang2025integrated}. On transmitting large-volume data, semantic communication selectively transmits mission-relevant data, thereby improving spectral efficiency under constrained conditions.

\textbf{Management subsystem}: Establishes an independent and resilient UAV control plane that is decoupled from mission-oriented data flows. Functional modules such as UAV Access Gateway Function (UAGF), UAV Access Control Function (UACF), UAV Registration Function (UARF), UAV Application Management Function (UAMF), and UAV Application Network Function (UANF) provide full lifecycle management for UAVs. This layer supports multi-standard access, cross-domain identity coordination, and rapid incident response, enabling regulatory oversight, service authentication, and control command dissemination in real time.

\textbf{Service subsystem}: Provides an orchestration layer for multi-role operations, integrating UAV task execution, supervisory control, and public safety services. With hierarchical deployment strategies spanning regional, provincial, and national levels, it supports customized, scenario-specific services while ensuring interoperability across UAV service domains. This enables dynamic route planning, adaptive task scheduling, and rapid risk mitigation across diverse LAE applications.

In summary, UTICN delivers a vertically integrated architecture that closes the loop from low-altitude situational sensing to decision-making and real-time control. Its cellular-native design forms a secure and intelligent digital foundation to ensure the safety, scalability, and serviceability of future low-altitude economic systems.

\section{Key Enabling Technologies} \label{sec-IV}

To realize UTICN, it is essential to overcome challenges across five core domains: sensing, positioning, communication, management, and service. This section introduces the key enabling technologies for UTICN.

\subsection{Multi-Domain Integrated Sensing}

Comprehensive sensing of low-altitude airspace requires coverage of both the physical space and the electromagnetic environment. As for the physical space, UTICN supports ISAC, which reuses spectrum and RF chains to support simultaneous communication and radar sensing. Using in-band coupling, radar symbols are embedded within control resource blocks. This design enables real-time tracking of small UAVs while maintaining high-throughput data links. Our field tests demonstrate 15 ms air-interface latency, 2 Gbps throughput, and 1.5 m range resolution, with successful multi-band cooperative detection of UAVs with radar cross-sections as low as 0.1 m$^2$.

Low-altitude airspace is also characterized by dense electromagnetic activity, where communication, navigation, and sensing signals often interfere with one another, especially in urban environments. To ensure robust situational awareness, UTICN integrates electromagnetic sensing across radar, optical, and radio frequency (RF) domains for seamless target tracking and interference localization.

To further enhance electromagnetic awareness, UTICN employs multi-domain analysis methods. These include automatic modulation classification to recognize UAV control links, radiation source fingerprinting across time-frequency-space domains, and blind protocol parsing to decode unknown signal formats. Finally, UTICN builds a unified multi-domain representation model using deep neural networks to capture emitter characteristics across multiple dimensions. Furthermore, dynamic control information is extracted to infer user behaviors and identify anomalies.

\subsection{Multi-Source Collaborative Positioning}

Accurate positioning in urban low-altitude environments is challenged by satellite signal occlusion from buildings, tunnels, and underground infrastructure. To ensure robust UAV tracking, UTICN adopts a multi-source fusion approach combining GNSS, ground-based differential corrections, pseudolite systems, and mobile network-based positioning, providing both active and passive positioning.

With recent breakthroughs in millimeter-wave communications and massive multiple-input multiple-output (MIMO), cellular carrier-phase positioning significantly improves ranging and angular resolution, offering centimeter-level positioning accuracy \cite{nikonowicz2024indoor}. Carrier-phase positioning extracts phase information from orthogonal frequency-division multiplexing (OFDM) multi-carrier signals. By resolving integer ambiguity and correlating phase with signal wavelength, UTICN achieves sub-decimeter accuracy. Using differential techniques across multiple ground stations, measurement noise is mitigated, providing higher positioning accuracy.

Depending on deployment, positioning can be calculated either actively, by aggregating measurements from ground base stations, or passively, with UAVs performing onboard localization using GNSS or real-time kinematic differential positioning signals. With improvements in onboard chips, carrier-phase positioning is feasible and scalable for UAV active positioning. To further enhance spatial resolution, UTICN implements multi-station active collaborative localization. ISAC-enabled base stations capture both self-transmitted echoes and echoes from neighboring stations. By leveraging inherent time-delay correlations, both active and passive signals are fused to reinforce positioning precision.

This approach yields three gains. First, coherent signal stacking significantly improves the signal-to-noise ratio. Second, passive signals synthesized from active measurements help suppress noise and enhance robustness. Third, sequential extension with phase-linked interpolation improves spatial resolution. This cross-station fusion architecture, integrating both passive and active positioning technologies, is capable of accurately tracking UAVs even in complex urban environments.

\subsection{Intelligent Machine Communication}

As the low-altitude economy scales, conventional UAVs relying on direct links over unlicensed bands (e.g., 2.4 GHz and 5.8 GHz) face increasing spectrum congestion and interference, particularly in urban environments. While proprietary data links remain useful, the proliferation of UAVs necessitates scalable, interference-resilient, and mission-critical connectivity solutions. UTICN addresses this by leveraging the cellular network infrastructure, dynamically adjusting antenna beam patterns to optimize coverage in near-ground airspace. The integration of 5G-A and ISAC technologies further strengthens air-ground connectivity, enabling wide-area, high-reliability links with situational awareness.

To balance performance and efficiency, UTICN employs 5G NR RedCap for mid-rate transmission tasks. RedCap, standardized in 3GPP Releases 17 and Releases 18 \cite{3gpp_tr_23_700_68}, reduces terminal complexity by constraining bandwidth, modulation schemes, and antenna count, resulting in up to 60\% cost and power savings while preserving key 5G capabilities such as low latency and edge intelligence. RedCap's compatibility with network slicing makes it ideal for UAV control and telemetry applications. Control UAV is logically isolated from mission payloads through network slicing and UPF-based routing, ensuring secure and differentiated service delivery.

Beyond air-ground links, UTICN introduces A2X technology to enable direct UAV-to-UAV communication, addressing challenges such as collision avoidance and swarm coordination. Built upon the evolution of Cellular Vehicle-to-Everything, A2X employs PC5 sidelink interfaces standardized in 3GPP Release 19 \cite{3gpp_ts_24577_v19_0_0}. It supports unicast, multicast, and broadcast over the 5.9 GHz, with NR-V2X protocol stack compatibility. A2X achieves sub-3 ms latency and over 1 km communication range in open environments, significantly outperforming legacy protocols such as Wi-Fi mesh or LoRa, which suffer from interference and limited throughput.

A2X also provides a unified communication framework for both air-to-air and air-to-ground interactions. UAVs can share identity, trajectory, and direct detect-and-avoid information with relevant stakeholders. This enables fine-grained trajectory deconfliction, formation control, and real-time situational awareness. Through integration with Uu interfaces and edge A2X service nodes, UAV controllers can maintain command-and-control links across network domains, ensuring continuous supervision and responsive control even in beyond-visual-line-of-sight scenarios.

By incorporating A2X into UTICN, UAVs gain not only autonomous negotiation capabilities but also network-assisted coordination via cellular base stations. This supports scalable UAV swarming, distributed perception, and multi-agent collaborative decision-making. Furthermore, A2X lays the foundation for a parallel, regulation-aware control network decoupled from mission data flows, enhancing resilience and enabling cooperative autonomy in dense, contested low-altitude environments.

For high-fidelity video transmission under bandwidth constraints, UTICN supports semantic communication techniques. Unlike traditional source-channel separation, semantic communication extracts and transmits task-relevant information based on shared knowledge between sender and receiver, significantly reducing data volume. By deploying pretrained semantic encoders/decoders, UAVs can achieve up to 60\%  data reduction in real-time video streaming without perceptual degradation, thus enhancing communication robustness in contested low-altitude environments. With advances in lightweight AI models and onboard acceleration, semantic communication emerges as a key enabler for future bandwidth-efficient UAV operations.

\subsection{Intelligent Concise Management}

To enable secure, scalable, and mission-critical UAV operations, UTICN introduces a dedicated control-data
separation framework that decouples the UAV control plane from mission data flows. This architecture is inspired by the Aeronautical Telecommunication Network used in civil aviation,  where UAVs operate outside traditional air traffic control systems and coexist with terrestrial public networks.

A specialized low-altitude UAV management network is constructed independently from the public internet, ensuring high-reliability services such as trajectory approval, cross-domain coordination, flight status monitoring, conflict avoidance, and emergency intervention. Functional modules, e.g., UAGF, UACF, UARF, UAMF, UANF, collaboratively support seamless UAV registration, identity synchronization, mobility management, and inter-domain handover. For example, UAGF provides a unified interface adaptation and seamless access across domains.

The architecture supports heterogeneous communication modalities, including 5G-A/A2X cellular interfaces, PC5-based direct communication, and proprietary UAV links.  To address regulatory constraints, the network supports localized control through hierarchical service domains, i.e., district, provincial, and national, ensuring governance compatibility with public safety, civil aviation, and military authorities.

To meet the real-time demands of UAV operations, the control plane integrates low-latency relay via edge UAGF, avoiding cloud backhaul and reducing control response times. It also supports audit trails and override commands such as forced hovering, return-to-home, or landing in case of emergencies, ensuring enforceable regulatory supervision.

Additionally, UTICN integrates mobile edge computing (MEC), UPF, network slicing, and QoS-aware orchestration to enable differentiated traffic routing for UAV control and mission payloads. UAVs dynamically access control and service providers, while UPF performs content-based steering toward MEC or clouds based on latency and bandwidth requirements. This architecture guarantees secure isolation and scalable service delivery across distributed UAV deployments.w

As UAV operations evolve toward swarm-based collaboration and AI-powered autonomy, UTICN supports hierarchical offloading strategies. Real-time control is kept on board, mission-critical computation is offloaded to MEC, and bulk tasks such as environmental modeling or multi-agent learning are routed to the cloud. A unified view of spectrum availability, computational load, and spatial-temporal airspace dynamics is maintained to support coordinated planning across physical, communication, and computing dimensions.

This holistic integration of control and network intelligence lays the groundwork for resilient, adaptive, and regulation-ready UAV airspace management, enabling high-density, low-altitude economic operations in future urban and peri-urban airspaces.
\begin{figure*}[t]
\centering
\includegraphics[width=0.8\linewidth]{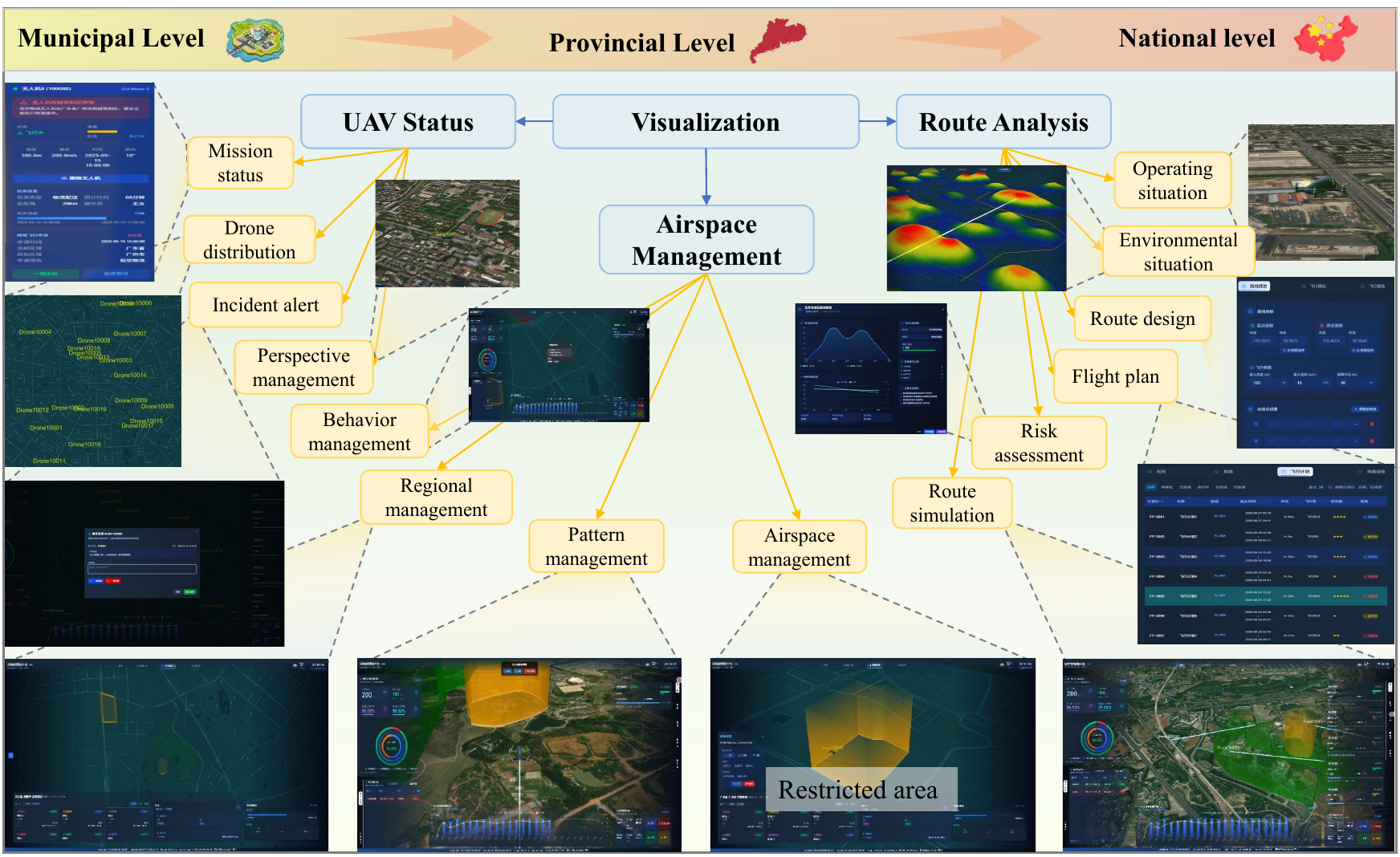}
\caption{Visualization of the function for the city-level LAE management platform.}
\label{fig:System}
\end{figure*}

\subsection{Swarm Intelligence-Based Situation Evolution}

The rapid increase in UAV density renders static, pre-planned routing strategies ineffective in managing large-scale LAE UAVs. Traditional traffic prediction methods are limited by the lack of historical data, while centralized multi-agent trajectory planning suffers from exponential complexity due to inter-UAV interactions in 3D airspace.

To address this, inspired by the traffic situation evolution in terrestrial transportation  \cite{luo2023alpharoute}, UTICN introduces a hierarchical situation evolution framework that integrates evolutionary game theory and Monte Carlo Tree Search (MCTS). Locally, UAVs use game-theoretic strategies to achieve user equilibrium, where each agent selects its optimal path based on local congestion and delay feedback. Globally, MCTS enables efficient exploration of multi-agent path combinations, leveraging pruning and risk evaluation to rapidly derive near-optimal swarm trajectories.

The framework captures the spatiotemporal coupling of low-altitude traffic by modeling airspace as a dynamic graph and adjusting routes based on real-time and predictive traffic states. This approach balances fairness and system efficiency, enabling minute-level adaptation and ensuring robust coordination across large UAV swarms in dynamic environments.

\section{Application Cases}
 \label{sec-v}
\subsection{Case 1: City-level LAE management platform}

A practical implementation of UTICN's management function is shown in Fig.~\ref{fig:System}. This system presents a pilot deployment in Guangzhou, China, serving as a benchmark for future low-altitude platforms. This system integrates key functionalities, including airspace delineation, UAV route planning, UAV monitoring, real-time alerting, situational awareness, and risk assessment.

Airspace and UAV operations are managed through fine-grained classification of controlled zones (e.g., restricted, temporary, or operational) and flight behaviors (e.g., unauthorized or compliant). It supports diverse routing modes such as fixed, dynamic, or free routes, and enables hierarchical monitoring of takeoff and landing sites from localized pads to full-scale urban ports.

The system provides a multi-level visualization interface (i.e., national, provincial, district), offering comprehensive insight into UAV distribution, mission status, alert conditions, and regional traffic dynamics. Users can interactively inspect and manage UAVs, flight paths, airspace conditions, and event responses.

The route analysis engine supports automatic and manual planning with simulation and risk assessment capabilities. Based on the situational analysis, flight anomalies are flagged, and dynamic risk maps are generated to inform airspace policy and route design.

The platform also features regional resource management, enabling visualization and dynamic adjustment of airspace, ground infrastructure, and operational boundaries. New zones can be added or adjusted with automatic alerts on their impact on existing flight routes.

Finally, the system integrates meteorological and geographic situational analysis, visualizing the effects of rainfall, wind, terrain, and signal quality. We adopt 3D grid models to assess flight suitability, offering situational support for the coordinated and safe operation of large UAV fleets.
\begin{figure*}
\centering
\includegraphics[width=0.9\linewidth]{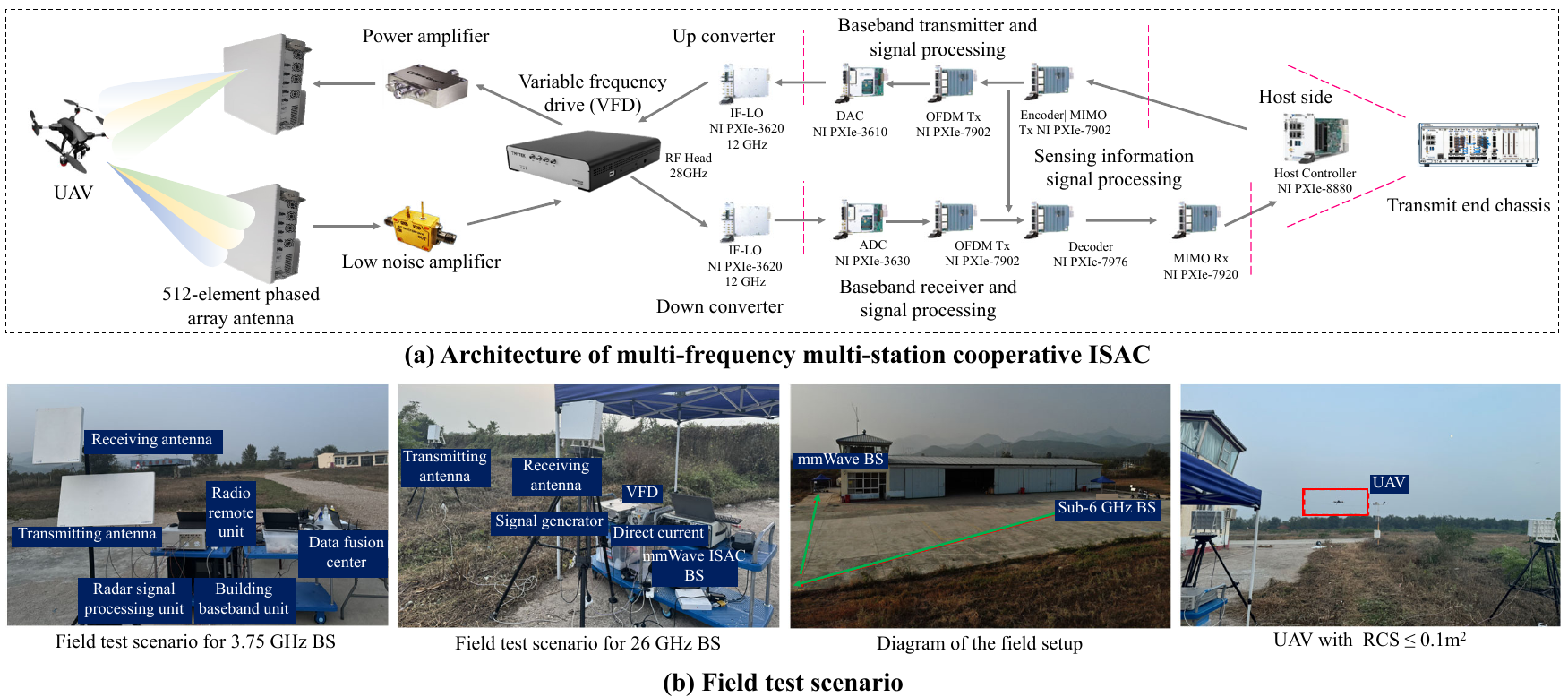}
\caption{Design and field test  of the prototype multi-frequency multi-station ISAC system.}
\label{fig:isac}
\end{figure*}

\subsection{Case 2: Multi-frequency Collaborative ISAC }

To validate UTICN's sensing and communication effectiveness in complex urban environments, we developed a prototype platform based on a multi-frequency ISAC-enabled collaborative sensing system. The system architecture (Fig. \ref{fig:isac}a) includes a baseband processor that maps user data into ISAC baseband signals, which are then upconverted to millimeter-wave or Sub-6 GHz frequencies and transmitted via high-power antennas. These waveforms simultaneously support radar-based object sensing and wireless communication.

Two ISAC base stations (BS) were deployed in our field test., i.e.,  Sub-6 GHz BS for wide-area sensing, working at 3.75 GHz, with 100 MHz bandwidth and 46 dBm effective isotropic radiated power (EIRP),  and millimeter-wave (mmWave) BS for close-range high-resolution detection,  working at 26 GHz, with 800 MHz bandwidth and 55 dBm EIRP. Both systems used OFDM waveforms with subcarrier-embedded radar samples for sensing-communication integration. The mmWave BS also employed massive MIMO and digital self-interference suppression to ensure precise signal reception. The sensing results of these two BSs are fused to form the observed trajectory of the target UAV.  More detailed parameter settings are shown in Table \ref{tab:params}.

\begin{table}[h]
\renewcommand{\arraystretch}{1.3}
\centering
\caption{Field test parameter settings.}
\label{tab:params}
\centering
\begin{tabular}{>{\raggedright\arraybackslash}p{0.09\textwidth}|>{\raggedright\arraybackslash}p{0.14\textwidth}|p{0.09\textwidth}}
\toprule
\textbf{Module} & \textbf{Parameter} & \textbf{ Value} \\
\hline
\multirow{6}{*}{mmWave BS} 
& Band & $26\,\text{GHz}$ \\
& Start/Stop frequency & $25.5$--$26.5\,\text{GHz}$ \\
& System bandwidth & $800\,\text{MHz}$ \\
& Antenna EIRP & $55\,\text{dBm}$ \\
& Symbol PRI & $0.2\,\text{ms}$ \\
& Subcarrier spacing & $75\,\text{kHz}$ \\
\hline
\multirow{6}{*}{Sub-6 GHz BS} 
& Band & $3.75\,\text{GHz}$ \\
& Start/Stop frequency & $3.7$--$3.8\,\text{GHz}$ \\
& System bandwidth & $100\,\text{MHz}$ \\
& Antenna EIRP & $46\,\text{dBm}$ \\
& Symbol PRI & $2.5\,\text{ms}$ \\
& Subcarrier spacing & $30\,\text{kHz}$ \\
\hline
\multicolumn{2}{l|}{UAV radar cross-sections} & $\leq 0.1\,\text{m}^2$ \\
\multicolumn{2}{l|}{Data fusion center refresh rate} & $1.5\,\text{s}$ \\
\bottomrule
\end{tabular}
\end{table}

In the experiment (Fig. \ref{fig:isac}b), a UAV with radar cross-sections less than 0.1 m$^2$ flew north to south, initially detected only by the Sub-6 GHz BS. Upon entering an overlapping coverage zone, both stations contributed to cooperative sensing. As the UAV moved eastward, the mmWave BS became the primary detector due to urban obstructions affecting the sub-6 GHz BS. A centralized processor fused multi-station data to generate a unified UAV trajectory.

\begin{figure}[t]
  \centering
  \begin{subfigure}[t]{0.75\linewidth}
    \centering
    \includegraphics[width=\linewidth]{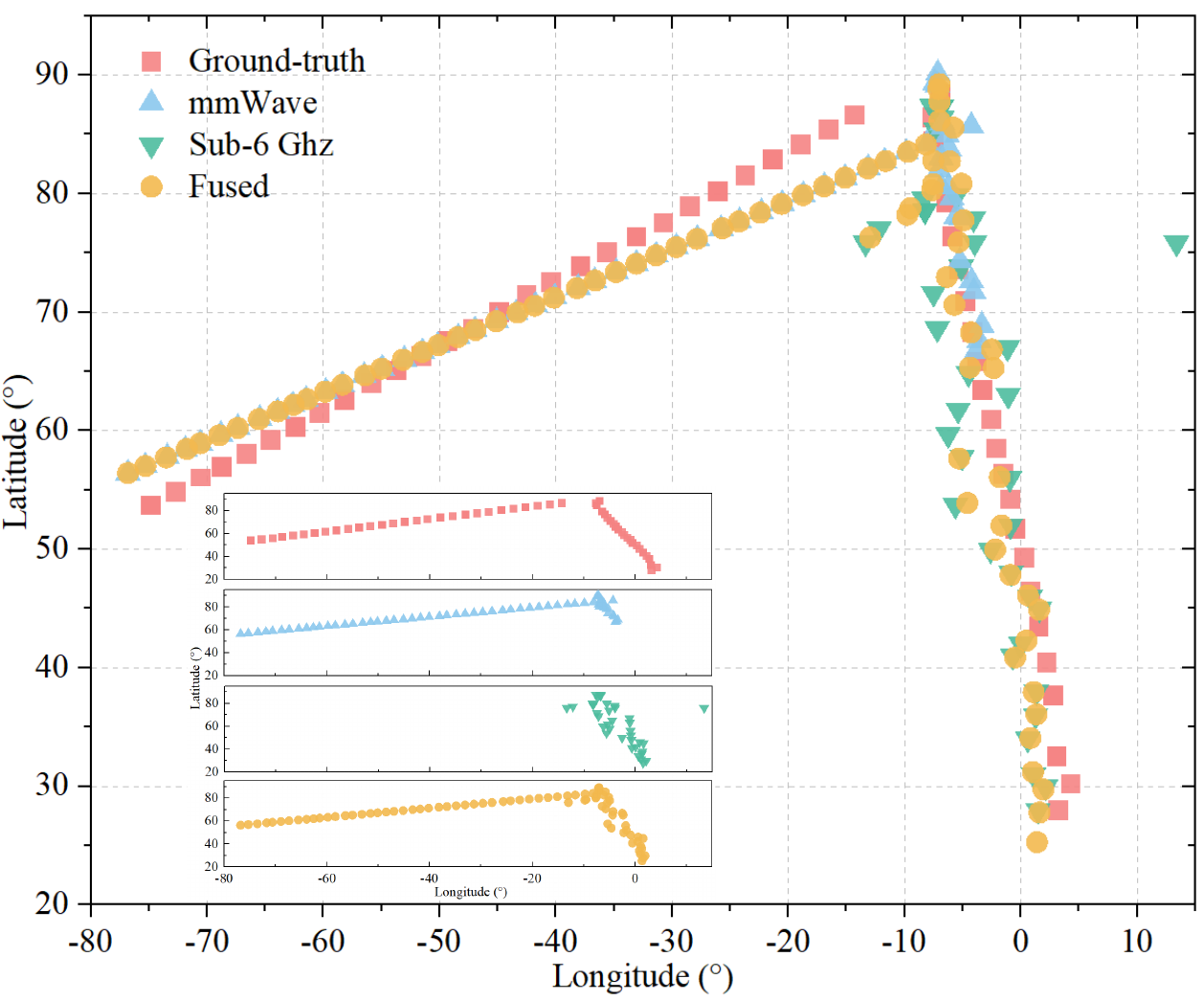}
    \caption{Trajectory of sensed UAV}
    \label{fig:fig-trajectory}
  \end{subfigure}

  \begin{subfigure}[t]{0.75\linewidth}
    \centering
    \includegraphics[width=\linewidth]{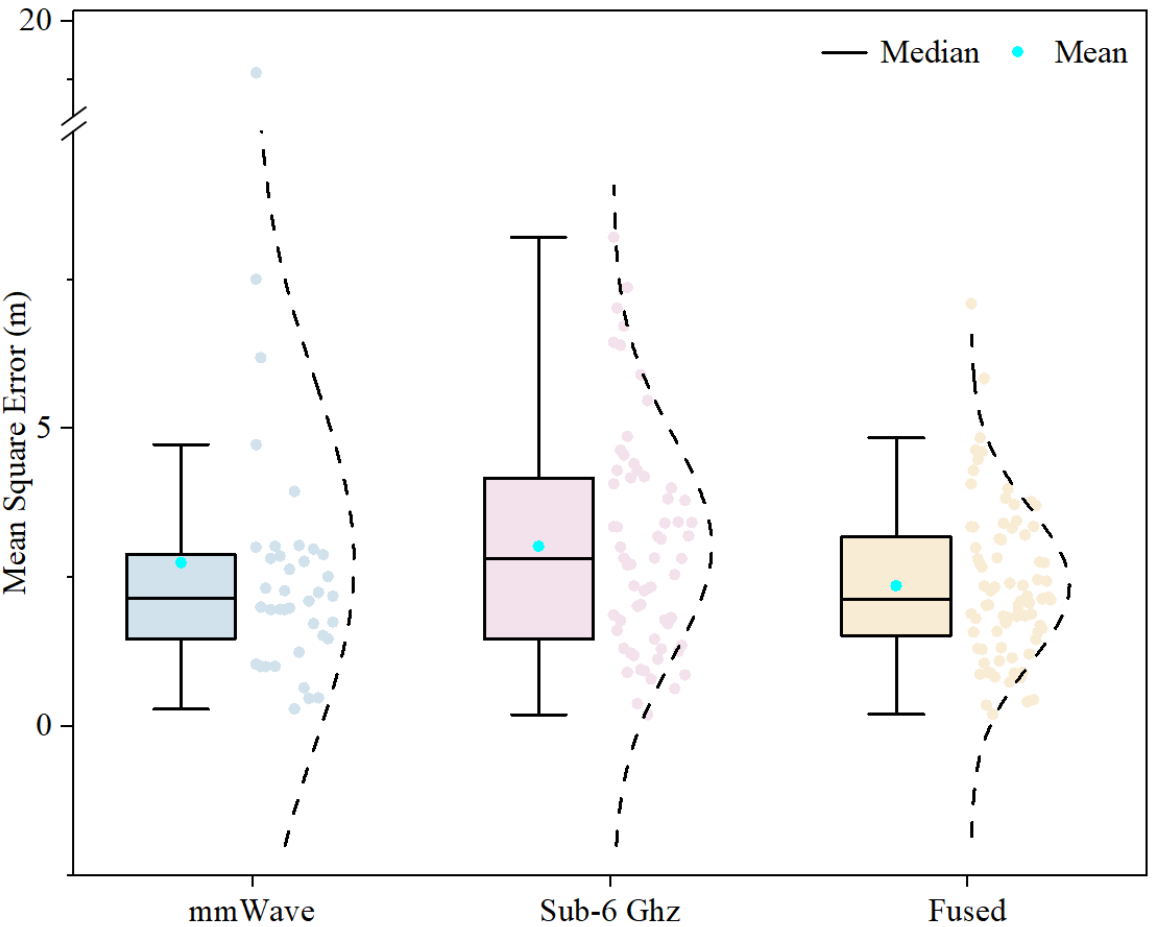}
    \caption{Sensing accuracy across different frequency bands}
    \label{fig:fig-mse}
  \end{subfigure}
  \caption{Performance of multi-frequency collaborative ISAC.}
  \label{fig:results}
\end{figure}

System performance was evaluated using ground-truth GPS trajectories. Positions with under-10-meter error were considered effective, following 3GPP TR22.837 standards. All 80 sampled trajectory points met this criterion, as shown in Fig. \ref{fig:fig-trajectory}. As illustrated in  Fig. \ref{fig:fig-mse}, Sub-6 GHz BS  showed a mean error of 2.5 meters, the mmWave BS  3.5 meters, and the fused estimate just 2.0 meters.  Furthermore, this experiment  demonstrates that fusion  over the mmWave BS 
 and Sub-6 GHz BS  enhanced continuity and tracking robustness under partial obstruction.

This case study confirms the feasibility and superior accuracy of multi-frequency, ISAC-enabled cooperative sensing in dense urban environments. The system offers valuable insights for applications in low-altitude traffic management, urban security, military reconnaissance, and disaster response.


\section{Conclusion} \label{sec-VI}

This article presents UTICN, a ubiquitous and Trusted Intelligent Cellular-native Network architecture purpose-built to support the intelligent, large-scale, ubiquitous, trusted, controllable, and operational development of LAE. By integrating multi-domain sensing, high-precision multi-source positioning, intelligent air-ground communication, control-data separated management, and swarm intelligence-driven situation evolution, UTICN offers a unified framework to address the pressing challenges of dynamic airspace management, UAV operation, and security management in complex low-altitude environments.

Through a modular cellular-native architecture and cross-domain technological integration, UTICN lays a robust foundation for future applications in urban airspace management, mission-critical UAV coordination, public security operations, and disaster response. As LAE systems continue to evolve, further research is needed in dynamic topology optimization, adaptive spectrum management, and multi-tier resource coordination. These efforts will enable seamless integration between terrestrial, aerial, and space-based networks, paving the way toward a fully connected, resilient, and context-aware low-altitude infrastructure ecosystem.

\normalsize{
		\bibliographystyle{IEEEtran}
		\bibliography{main}
}

\end{document}